\documentclass[a4paper]{jpconf}
\usepackage{feynmp,feynmp-sup,epsfig,amssymb,cite}

\def\Gdashed{
\parbox{18\unitlength}{
\begin{fmfgraph*}(18,10)
\fmfleft{l}
\fmfright{r}
\fmf{dbl_plain,label=$i\hspace*{14\unitlength}j$,l.s=left}{r,l}
\end{fmfgraph*}
}}
\def\Pdashed#1#2{
\parbox{18\unitlength}{
\centerline{\begin{fmfgraph*}(18,10)
\fmfleft{l}
\fmfright{r}
\fmf{dashes_arrow,label=$#1\hspace*{14\unitlength}#2$,l.s=left}{r,l}
\end{fmfgraph*}}
}}
\def\FBox#1#2#3{ 
\parbox{19\unitlength}{
\begin{fmfgraph*}(19,26)
\fmfpen{thin}
\fmfleft{lb,lt}
\fmfright{rb,rt}
\fmf{phantom,tension=2}{rb,prb}
\fmf{phantom,tension=2}{lb,plb}
\fmf{phantom,tension=2}{rt,prt}
\fmf{phantom,tension=2}{lt,plt}
\fmfpoly{filled=30,label=#3}{prb,prt,plt,plb}
\fmfdot{plb,prb}
\fmflabel{#1}{plb}
\fmflabel{#2}{prb}
\end{fmfgraph*}
}}
\def\FBoxL{
\parbox{22\unitlength}{$ $\hfill
\begin{fmfgraph*}(19,26)
\fmfpen{thin}
\fmfleft{lb,lt}
\fmfright{rb,rt}
\fmf{phantom,tension=2}{rb,prb}
\fmf{phantom,tension=2}{lb,plb}
\fmf{phantom,tension=2}{rt,prt}
\fmf{phantom,tension=2}{lt,plt}
\fmfpoly{filled=30,label=$M'$}{prb,prt,plt,plb}
\fmffreeze
\fmf{dashes_arrow,left=0.8}{plb,plt}
\fmf{dbl_plain,left=0.8}{plt,prt}
\fmfdot{plb,prb}
\fmflabel{$1$}{plb}
\fmflabel{$2$}{prb}
\fmflabel{$3$}{plt}
\fmflabel{$4$}{prt}
\end{fmfgraph*}
}}
\def\FBoxR{
\parbox{22\unitlength}{
\begin{fmfgraph*}(19,26)
\fmfpen{thin}
\fmfleft{lb,lt}
\fmfright{rb,rt}
\fmf{phantom,tension=2}{rb,prb}
\fmf{phantom,tension=2}{lb,plb}
\fmf{phantom,tension=2}{rt,prt}
\fmf{phantom,tension=2}{lt,plt}
\fmfpoly{filled=30,label=$M'$}{prb,prt,plt,plb}
\fmffreeze
\fmf{dashes_arrow,right=0.8}{prb,prt}
\fmf{dbl_plain,left=0.8}{plt,prt}
\fmfdot{plb,prb}
\fmflabel{$1$}{plb}
\fmflabel{$2$}{prb}
\fmflabel{$3$}{plt}
\fmflabel{$4$}{prt}
\end{fmfgraph*}\hfill$ $
}}
\def\Ploop{\parbox{11\unitlength}{
\begin{fmfgraph*}(10,10)
\fmfpen{thin}
\fmftop{t}
\fmfleft{l}
\fmfbottom{b}
\fmf{dashes_arrow,left=1.0,tension=1}{t,b}
\fmf{dashes,left=1.0,tension=1}{b,t}
\fmfdot{l}
\end{fmfgraph*}
}}
\def\DRhomb#1#2{ \parbox{33\unitlength}
  {\centerline{\begin{fmfgraph**}(30,15,11) \fmfright{r} \fmfleft{l}
        \fmfforce{0.5w,0.5h}{m} \fmfpoly{filled=30,label=#1}{m,t1,l,b1}
        \fmfpoly{filled=30,label=#2}{m,b2,r,t2} \fmfdot{r,m,l}
\end{fmfgraph**}}}
}
\def\DPoisson{\parbox{33\unitlength}
{\centerline{\begin{fmfgraph**}(30,15,11)
\fmfright{r}
\fmfleft{l}
\fmfforce{0.5w,0.5h}{m}
\fmfpoly{filled=30,label=$A$}{m,t1,l,b1}
\fmfpoly{filled=30,label=$\partial_X B$}{m,b2,r,t2}
\fmf{dashes_arrow,left=1.3}{l,m}
\fmfdot{r,m,l}
\end{fmfgraph**}}}
+
\parbox{33\unitlength}
{\centerline{\begin{fmfgraph**}(30,15,11)
\fmfright{r}
\fmfleft{l}
\fmfforce{0.5w,0.5h}{m}
\fmfforce{0.25w,0.5h}{ll}
\fmfforce{0.75w,0.5h}{rr}
\fmfpoly{filled=30,l.d=0,label=$\partial_X A$}{m,t1,l,b1}
\fmfpoly{filled=30,label=$B$}{m,b2,r,t2}
\fmf{dashes_arrow,right=1.3}{r,m}
\fmfdot{r,m,l}
\end{fmfgraph**}}}
}

\def\scr#1{{\mathrm{ #1}}}
\def\Gr{G}\def\Se{\Sigma}\def\ii{{\rm i}}\def\di{{\rm d}}\def\MP{\mp}
\def\cal{\mathcal}
\def\intp{\int\frac{\di^4 p}{(2\pi)^4}}
\def\oint{\int_{\cal C}}
\def\E{{\cal E}}
\renewcommand{\Re}{{\mathrm{Re}}}
\renewcommand{\Im}{{\mathrm{Im}}}

\def\vec#1{\mbox{\boldmath $#1$}}
                
\newcommand{\Pbr}[1]{\left\{#1\right\}}

\begin{document}

\title{Dynamics of Resonances in Strongly Interacting Systems}

\author{J. Knoll$^1$, F. Riek$^1$,
Yu. B. Ivanov$^{1,2}$ and D.N. Voskresensky$^{1,3}$}
\address{$^1$ Gesellschaft f\"ur Schwerionenforschung mbH, Planckstr. 1\\
64291 Darmstadt, Germany}
\address{$^2$ Kurchatov Institute, Kurchatov sq. 1, Moscow 123182,
Russia}
\address{$^3$ Moscow Institute for Physics and Engineering, 
Kashirskoe sh. 31, Moscow 115409, Russia}
\ead{J.Knoll@gsi.de, F.Riek@gsi.de, Y.Ivanov@gsi.de, D.Voskresensky@gsi.de}

\begin{abstract} The effects of the propagation of particles which
  have a finite life-time and an according broad distribution in their
  mass spectrum are discussed in the context of a transport
  descriptions. In the first part some example cases of mesonic modes
  in nuclear matter at finite densities and temperatures are presented.
  These equilibrium calculations illustrate the dynamical range of
  spectral distributions to be adequately covered by non-equilibrium
  description of the dynamics of two nuclei colliding at high
  energies.  The second part addresses the problem of transport
  descriptions which properly account for the damping width of the
  particles. A systematic and general gradient approximation is
  presented in the form of diagrammatic rules which permit to derive a
  self-consistent transport scheme from the Kadanoff--Baym equation.
  The scheme is conserving and thermodynamically consistent provided
  the self-energies are obtained within the $\Phi$-derivable
  two-particle irreducible (2PI) method of Baym. The merits, the
  limitations and partial cures of the limitations of this transport
  scheme are discussed in detail.\\[-1cm]
\end{abstract}

\section{Scope of high energy nuclear collisions}

Two atomic nuclei collide at high relative energies. During a time
span of less than 10$^{-22}$ sec. a highly compressed and heated
interaction zone is formed with matter densities $\rho$ that
exceed several times the nuclear saturation density $\rho_0$ of 0.16
nucleon per cubic femto-meter (fm$^3$) and temperatures $T$ expressed
in energy units beyond 100 MeV, cf. Fig. 1, below \cite{3FL-Hydro}. In
daily life units this corresponds to $\rho\ge 10^{15}$ g/cm$^3$ and
$T\ge 10^{12}$ K and thus to the highest matter densities and
temperatures ever realized in laboratory experiments.  Such matter
conditions prevailed during the first few micro-seconds of the early
universe, occur in super nova explosions and their remnants, the
neutron stars at somewhat lower temperatures.  This form of matter can
only be investigated in the laboratory through nuclear collisions
where one collects the reaction products emitted from the highly
excited collision zone in sophisticated multi-particle detector
arrangements. For a recent and comprehensive experimental overview see
e.g. ref.  \cite{AAPBM}. In such experiments one distinguishes
strongly interacting reaction products, such as hadrons, essentially
nucleons and pions but also hadrons with other flavors like
strangeness or charm from solely electro-magnetically interacting
probes such as photons or lepton pairs (electrons-positrons or
di-muons).  While the former can only give a clear view back to the
situation where the particles were essentially frozen out, i.e. the
strong interactions cease, the latter are classified as penetrating
probes.  Once created these particles pass the collision zone
essentially without further interaction. Such penetrating probes
permit to look deeply into the interaction zone and are thus
interesting as messengers from the compressed and heated matter.
\begin{center}
{\includegraphics[width=13cm]{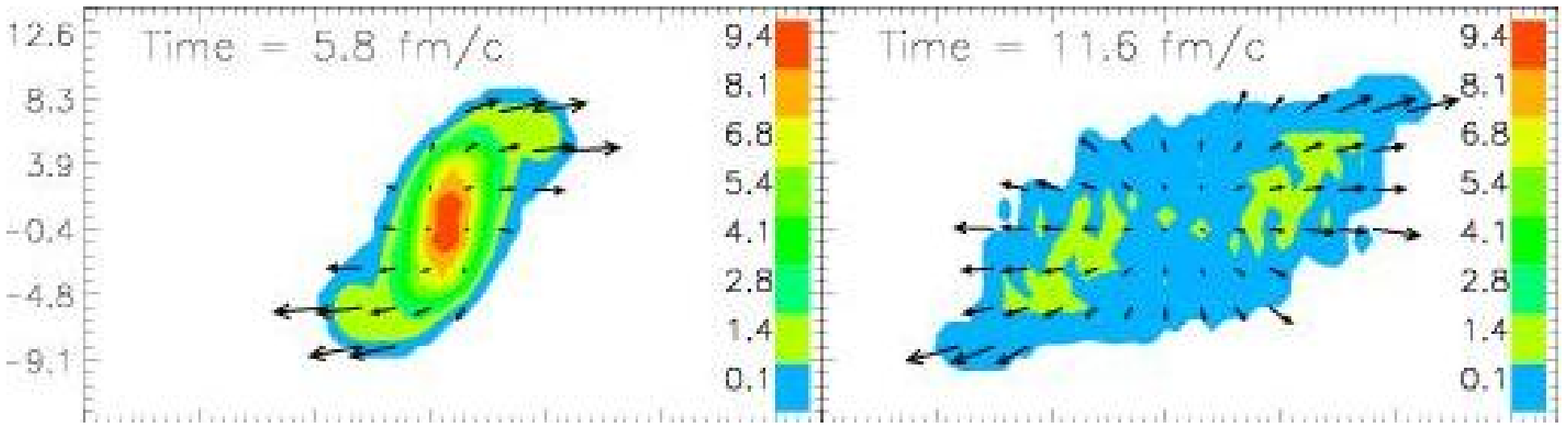}}\\[5mm]
  \parbox{12cm}{
{\bf Figure 1.} {Collision of two gold nuclei at
    a bombarding energy per nucleon of 160 GeV. Contour plots of the
    densities in units of $\rho_0$ at two time steps, resulting from a
    three-fluid hydrodynamical model
    \cite{3FL-Hydro}\label{3FL-Hydro}.\\ $ $}}
\end{center}
The observation of di-leptons has received much attention over the
last decade due to the observed enhancement of di-lepton pairs in the
invariant mass region below the vector-meson masses \cite{CERES,NA50}
and the observed suppression of $J/\Psi$-mesons in central nuclear
collisions compared to proton-proton collisions. With the recently
completed HADES di-electron spectrometer GSI continues this research
at lower bombarding energies. Considerable progress in mass resolution
achieved by new pixel detectors for high energy muons has enabled new
experimental results with unprecedented mass resolution
\cite{NA60,Damjanovic}.

The endeavor to investigate such nuclear collisions in experiments at
GSI, CERN or at the RHIC-collider at the Brookhaven laboratory (BNL)
strongly relies on adequate theoretical descriptions of the dynamics
of such reactions. While a fully quantum mechanical or even field
theoretical description is off reach, simplified approaches are
considered that cover different dynamical ranges of applicability.
With focus on the equation of state (EoS) of the interacting matter
hydrodynamical models were investigated. Since a full instantaneous
stopping to local equilibrium is not compatible with experimental
observations, various kinds of multi-fluid models were considered to
relax these constraints and to allow for a partial transparency
between the different fluid components \cite {3FL-Hydro}, cf. the
example given in Fig. 1. Even further generalizations treated the
abundances of the different constituents by non-equilibrium rate
equations, while only the kinetic degrees of freedom and the pressure
were considered to be in local equilibrium \cite{flavor}. These
approaches permit to address the issues of phase transitions e.g.
between the hadronic and the quark-gluon plasma (QGP) phase and the
underlying EoS.

At the microscopic level kinetic transport descriptions are in use.
They describe the constituents of the matter as classical particles
subjected to Newtonian forces and to stochastic collisions among the
particles in terms of generalized Boltzmann equations which adopt mean
fields and the statistics (Fermi-Dirac/ Bose-Einstein) of the
particles. Thereby the collision term is essentially taken from
binary-collision cross-sections. The relevant degrees of freedom may
significantly change during the collision process.  Due to the
collisions among the incident nucleons the latter do not only change
their momenta but rather also may get excited to baryon resonances,
further more mesons and mesonic resonances can be created and at
sufficiently high energy density the sub-hadronic degrees of freedom,
the quarks and gluons, become liberated forming a QGP.

In such collisions one is confronted with the fact that the dynamics
has to include particles like the Delta isobar or e.g. rho-meson
resonances which have life-times of less than 2 fm/c already in the
vacuum. The equivalently damping widths thus exceed values of 100 MeV.
Also the collisional damping rates deduced from presently used
transport codes are of the same order, whereas typical mean
temperatures range between 50 to 150 MeV depending on beam energy.
Thus, the damping widths of most of the constituents in the system can
by no means be treated as a perturbation.  As a consequence the mass
spectrum of the particles in the dense matter is no longer
characterized by a sharp delta-function but rather acquires a width
due to collisions and decays. One thus comes to a picture which
unifies {\em resonances} that have already a width in vacuum due to
decay modes with the ''states'' of particles in dense matter, that
obtain a width due to collisions (collisional broadening).

The theoretical concepts for a proper many body description in terms
of a real time non-equilibrium field theory were already devised by
Schwinger~\cite{Schw}, Kadanoff and Baym~\cite{KB}, and
Keldysh~\cite{Keld64} in the early sixties, extensions to relativistic
plasmas followed by Bezzerides and DuBois~\cite{Bez}.  First
investigations of the quantum effects on the Boltzmann collision term
were given by Danielewicz~\cite{D}, the principal conceptual problems
on the level of quantum field theory were investigated by
Landsmann~\cite{Landsmann}, while applications which seriously include
the finite width of the particles in transport descriptions were
carried out only in recent times, e.g.
refs.~\cite{D,DB,BM,HFN,PH,QH,Weinhold,KV,Leupold,CJ}.  For
resonances, e.g. the Delta resonance, it was natural to consider broad
mass distributions and ad hoc recipes have been invented to include
this in transport simulation models.  However, many of these recipes
were not correct as they violated some basic principles like detailed
balance (see discussion in ref.~\cite{DB}), and the description of
resonances in dense matter has to be improved~\cite
{Weinhold,KV,IKV99,IKV00,Leupold,CJ,IKHV,IKV03}. A further conceptual
problem in self-consistent resummation schemes concerns the proper
renormalization of the (hidden) divergent loops in the
scheme. In this question significant progress has recently been
achieved \cite{vanHeesKnoll,Reinosa03,Reinosa05}.

In this contribution the consequences of the propagation of particles
with short life-times are discussed. First we investigate at the
example of nuclear matter at various temperatures a strongly
interacting system in equilibrium. In particular the spectral
properties of pionic modes in matter are explored. These pionic modes
couple to vector mesons which through their electromagnetic decay into
di-leptons (pairs of electron-positron or muons) are considered as
special messengers of the dense matter phase. These example cases
illustrate the properties of particles with broad damping width in
dense matter environment.  In the final part we discuss how particles
with such broad mass-width can be described consistently within a
transport theoretical picture.

We are going to argue that the Kadanoff--Baym equations in the first
gradient approximation together with the $\Phi$-functional method of
Baym~\cite{Baym} provide a proper frame for kinetic description of
systems of particles with a broad mass-width. To this end, we discuss
relevant problems concerning charge and energy--momentum conservation,
thermodynamic consistency, memory effects in the collision term and
the growth of entropy in specific cases\cite{IKV99}.

\section{Equilibrium properties of particles and resonances in
  dense nuclear matter}
In nuclear physics quite a broad variety of concepts and methods are
in use depending on the particular question raised. This is due to the
non-perturbative nature of the interaction. For nuclear structure
commonly non-relativistic concepts are used, nowadays with so-called
realistic two-body interactions, supplemented with a three-body
correction. There the essential question lies in the proper treatment
of the short range and tensor correlations \cite{FMD,UCOM} with unifying
prospects in the low-energy limit of a derived effective interaction
\cite {V_low_k}.  In such approaches the exchanged bosons which
furnish the forces only implicitly enter the dynamics through the
static two-body potentials.

For the here presented treatment with focus on high energy excitations
the exchange bosons become dynamical. Therefore we employ an effective
field theory picture with nucleons, and the Delta-isobar resonance as
the main baryonic matter fields interacting via pion exchange for the
long-range part of the interaction. This we supplement by a repulsive
Migdal-type two-body interaction of zero range in order to provide
proper saturation and other properties of nuclei.  Interested in
penetrating probes we extend the model to include the lightest vector
mesons in the second part of this section. These vector mesons are
discussed as important messengers about the dense nuclear environment
produced in nuclear collisions, since they can be observed through
their electromagnetic decay into electron-positron or muon pairs.

\subsection{Pions and baryonic resonances}\label{subsect-piNDelta}
For the description of the nuclear matter we consider the nucleon and
the $\Delta$(1232)-resonance as the main baryonic degrees of freedom.
The pion is the lightest strongly interacting boson. Thereby it is
commonly considered as a Goldstone boson related to the approximate
chiral symmetry which would exist for massless pions. In lowest
approximation in the pion-energy the underlying symmetry leads to
$p$-wave couplings to the baryonic pseudo-vector currents together
with a minor $s$-wave term. The latter we ignore. Thus one has a
standard 3-vertex which couples the baryonic current to the pion. For
the energies involved we approximate the Dirac spinor structure of the
baryons by the non-relativistic limit in no sea approximation
(ignoring anti-baryons), while the kinematics is kept in the
relativistic form.  The explicit analytic expressions and further
details can be found in ref. \cite{RK},see earlier work in this
context \cite{KorpaMalfliet,Lutz03}. To this end the pion
self-energy is given by simple particle-hole and Delta-hole loops,
while the baryon self-energies involve a single rainbow type loop
which includes one pion line.

The self-consistent coupled Dyson scheme includes all terms resulting
from a RPA treatment of the one-pion exchange interaction between the
baryons.  The latter accounts for the long-range attraction among the
nucleons.  Besides this one needs a repulsive interaction of short range
between the nucleons in order to stabilize the matter. This
interaction is commonly approximated by a zero range form with a 4-baryon
vertex, initially introduced by Migdal\cite{migdal}. Iterating the
corresponding nucleon-hole and Delta-hole loop insertions to all orders
the retarded pion self-energy
\begin{equation}
        \Pi^R_{\pi}=(\textbf{q}^2)^2\frac{\Pi^R_{\pi\; NN^{-1}}
          +\Pi^R_{\pi\; \Delta h}-(g_{11}-2g_{12}
          +g_{22})\Pi^R_{\pi\;NN^{-1}}\Pi^R_{\pi\;\Delta h}}
        {(\textbf{q}^2-g_{11}\Pi^R_{\pi\;NN^{-1}})(\textbf{q}^2
          -g_{22}\Pi^R_{\pi\;\Delta h})
          -g_{12}^2\Pi^R_{\pi\;NN^{-1}}\Pi^R_{\pi\;\Delta h}}F(\textbf{q})
        \label{Migdal1}
\end{equation}
 can still be obtained in closed form
\cite{DmitrievSuzuki,Urban}. Here and in the following the
boson self-energies are denoted by $\Pi$. The $g_{ik}$ denote the
Migdal parameters and $F(q)$ is a pion-baryon form-factor that affects
the high energy-momentum behavior of the pion. The particle-hole
loops are given by
\begin{fmffile}{fmfKiel}
\begin{displaymath}
\Pi^R_{\pi\;NN^{-1}}=
\parbox{27\unitlength}{
\begin{fmfgraph*}(25,15)
          \fmfpen{2}
          \fmfforce{0.1w,0.5h}{i}
          \fmfforce{1w,0.5h}{o}
          \fmfforce{0.225w,0.5h}{v1}
          \fmfforce{0.875w,0.5h}{v2}
          \fmf{dashes,fore=red}{i,v1}
          \fmf{dashes,fore=red}{v2,o}
          \fmf{fermion,left=0.7,tension=.5,fore=blue}{v1,v2}
          \fmf{fermion,left=0.7,tension=.5,fore=blue}{v2,v1}
\end{fmfgraph*}},\qquad
\Pi^R_{\pi\;\Delta h}=
\parbox{27\unitlength}{
\begin{fmfgraph*}(25,15)
          \fmfpen{2}
          \fmfforce{0.1w,0.5h}{i}
          \fmfforce{1w,0.5h}{o}
          \fmfforce{0.225w,0.5h}{v1}
          \fmfforce{0.875w,0.5h}{v2}
          \fmf{dashes,fore=red}{i,v1}
          \fmf{dashes,fore=red}{v2,o}
          \fmf{dbl_plain_arrow,left=0.7,tension=.5,fore=red+blue}{v1,v2}
          \fmf{fermion,left=0.7,tension=.5,fore=blue}{v2,v1}
\end{fmfgraph*}}.
\end{displaymath}
Likewise the nucleon and $\Delta$-resonance self-energies are given by
\begin{eqnarray}\nonumber
\Sigma^R_{N}&=&
\parbox{27\unitlength}{
\begin{fmfgraph*}(25,15)
          \fmfpen{2}
          \fmfforce{0.1w,0.5h}{i}
          \fmfforce{1w,0.5h}{o}
          \fmfforce{0.225w,0.5h}{v1}
          \fmfforce{0.875w,0.5h}{v2}
          \fmf{plain,fore=blue}{i,v1}
          \fmf{plain,fore=blue}{v2,o}
          \fmf{fermion,left=0.7,tension=.5,fore=blue}{v1,v2}
          \fmf{dashes,left=0.7,tension=.5,fore=red}{v2,v1}
\end{fmfgraph*}}
+
\parbox{27\unitlength}{
\begin{fmfgraph*}(25,15)
          \fmfpen{2}
          \fmfforce{0.1w,0.5h}{i}
          \fmfforce{1w,0.5h}{o}
          \fmfforce{0.225w,0.5h}{v1}
          \fmfforce{0.875w,0.5h}{v2}
          \fmf{plain,fore=blue}{i,v1}
          \fmf{plain,fore=blue}{v2,o}
          \fmf{dbl_plain_arrow,left=0.7,tension=.5,fore=red+blue}{v1,v2}
          \fmf{dashes,left=0.7,tension=.5,fore=red}{v2,v1}
\end{fmfgraph*}
},\qquad
\Sigma^R_{\Delta}=
\parbox{27\unitlength}{
\begin{fmfgraph*}(25,15)
          \fmfpen{2}
          \fmfforce{0.1w,0.5h}{i}
          \fmfforce{1w,0.5h}{o}
          \fmfforce{0.225w,0.5h}{v1}
          \fmfforce{0.875w,0.5h}{v2}
          \fmf{dbl_plain,fore=red+blue}{i,v1}
          \fmf{dbl_plain,fore=red+blue}{v2,o}
          \fmf{fermion,left=0.7,tension=.5,fore=blue}{v1,v2}
          \fmf{dashes,left=0.7,tension=.5,fore=red}{v2,v1}
\end{fmfgraph*}}.
\end{eqnarray}

The corresponding self-consistent results for the pion spectral
function are shown in the Figs. 2 and 3.  In the spectral function of
the pion we see the effect of the coupling to the different baryonic
excitations. The on-shell pole broadens and is shifted towards lower
energies due to the net-attraction of the interaction in this
kinematical region. The pion strongly mixes with the $\Delta N^{-1}$
component. The latter leads to the shoulders around the peak. This
component appears in the time-like region for low momenta, while it
traverses to the space-like region for higher momenta. At low energies
an entirely space-like pion component appears due to $NN^{-1}$
excitations. From the kinematics it is clear that this component stays
in the space-like region for all momenta.  Such space-like pion modes
have to be interpreted as a scattering process between the baryons in
the matter mediated by the exchange of pions. Due to this component,
the pion spectral function has non-vanishing strength at all energies,
such that in all processes where pions are involved all thresholds
disappear. The short-range Migdal repulsion prevents pion condensation
to occur at nuclear saturation density.  The form of the spectral
function turns out to be sensitive to values of the Migdal parameters
and the pion baryon form-factor $F(q)$ employed in the model. Due to
this form-factor, which is fitted to the $\pi N$-scattering data the
$\Delta$-hole component appears only as a slight shoulder above the
peak. For the resolution used in the calculations the well known zero
sound component at zero $T$ is invisible.

For higher temperatures all structures are broadened. It is clear that
such structures can no longer be described by a quasi-particle
picture. Rather one has to take due account of the fully dressed
spectral distributions of the particles in the medium.
 \hspace*{3mm}       \begin{minipage}[t]{.45\linewidth}
 \centerline{\epsfig{file=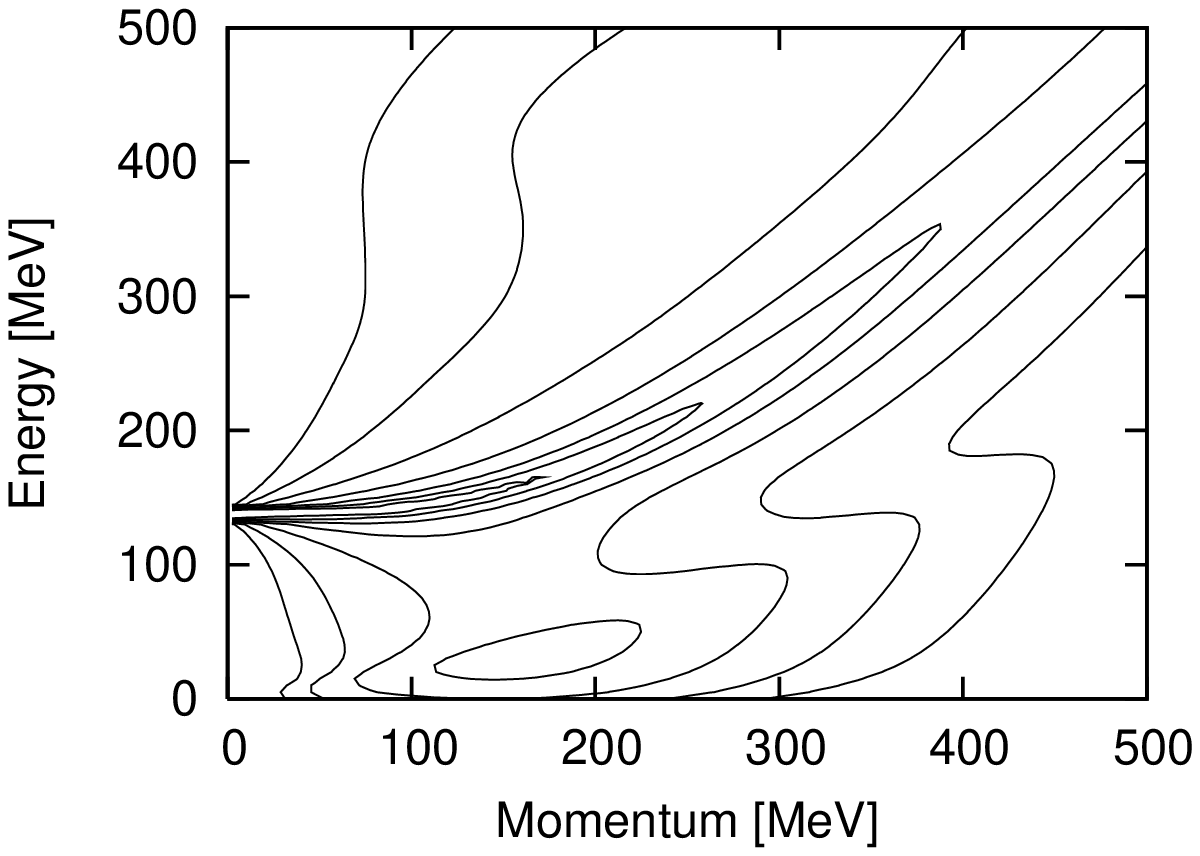,height=5. cm,width=4. cm,angle=-90}}
 $ $\\[-4mm] 
  \centerline{\epsfig{file=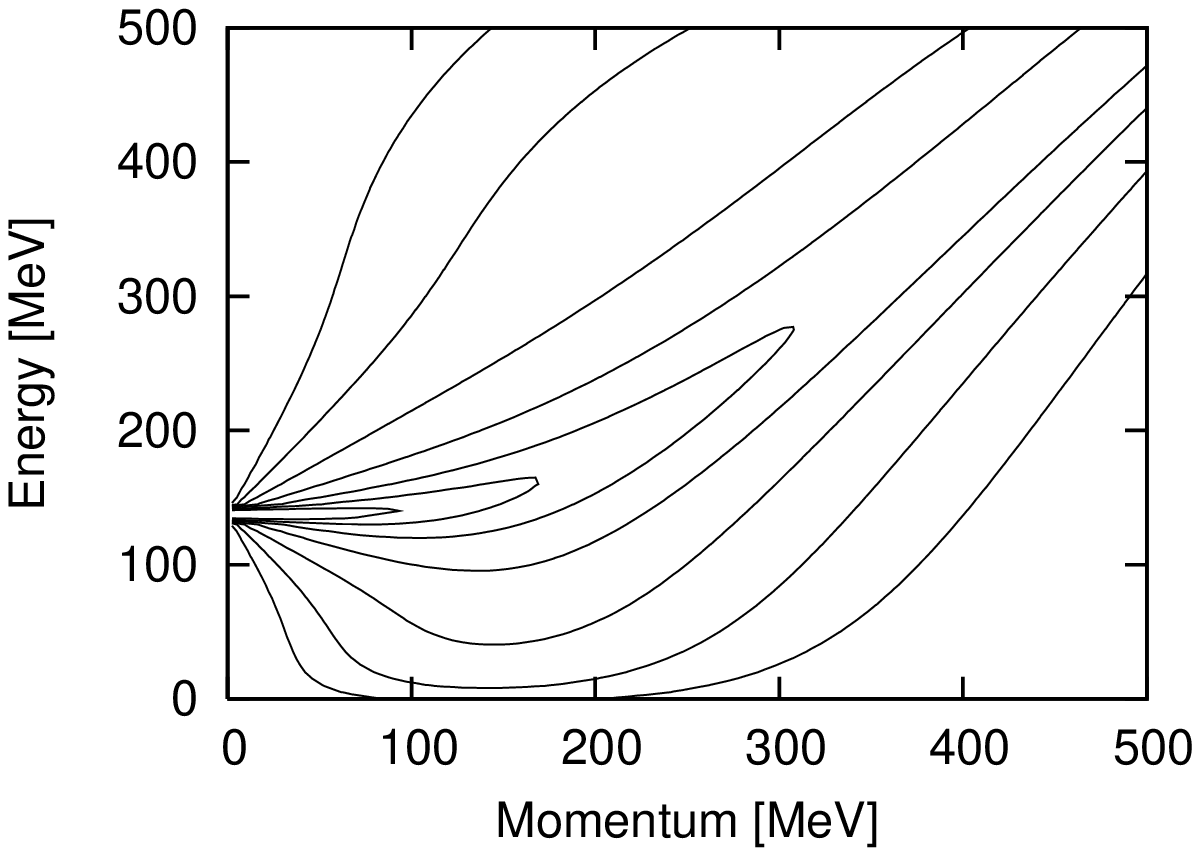,height=5 cm,width=4
      cm,angle=-90}} $ $\\ 
{
{\bf Figure 2.} {Contour plot of the logarithm of the pion
  spectral function $ 
  A_\pi$ at T=0 and 120 MeV (upper/lower panel) and $\rho
  =\rho_0$. The line spacing 
  covers half a decade.\\ $ $}\label{contour-pi}}
         \end{minipage}\hfill
       \begin{minipage}[t]{.45\linewidth}
 \centerline{\epsfig{file=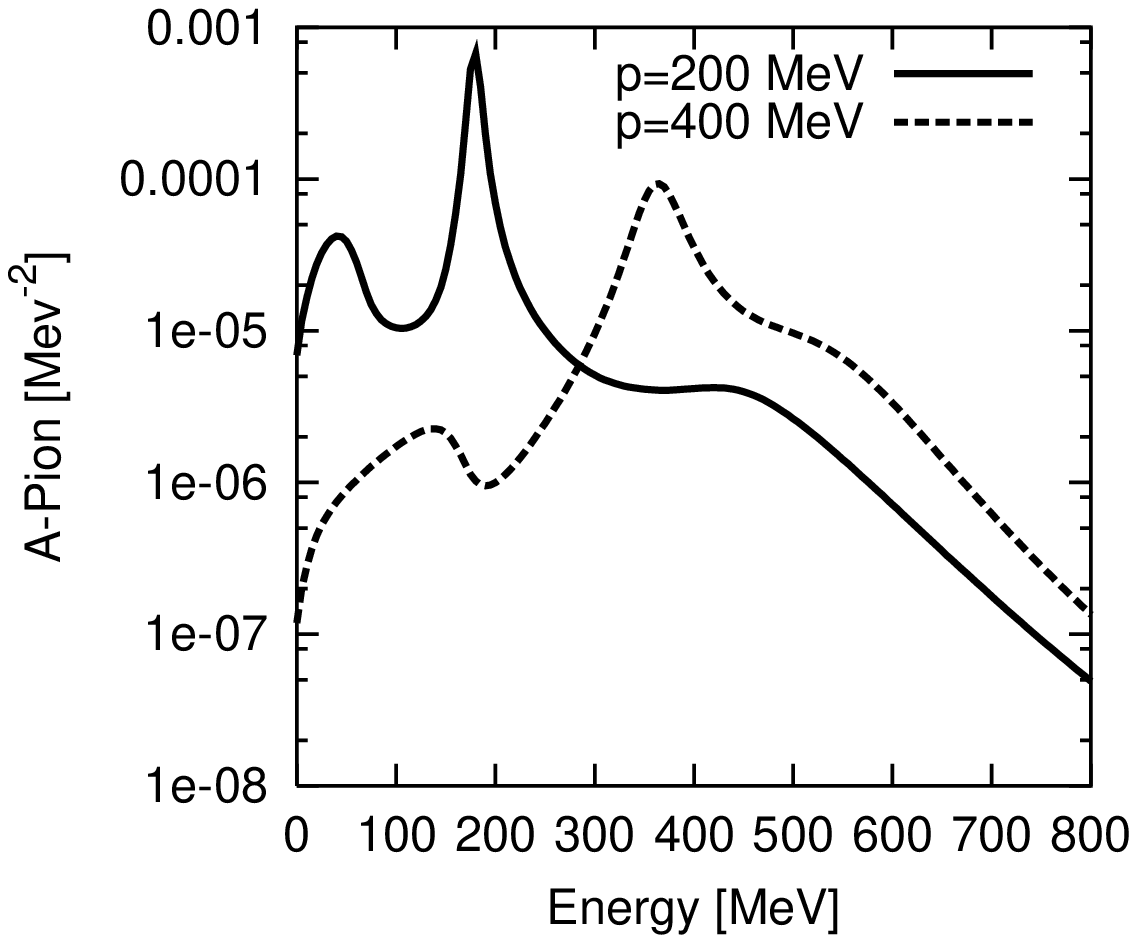,height=5. cm,width=4.
            cm,angle=-90}}$ $\\[-4mm]
   \centerline{\epsfig{file=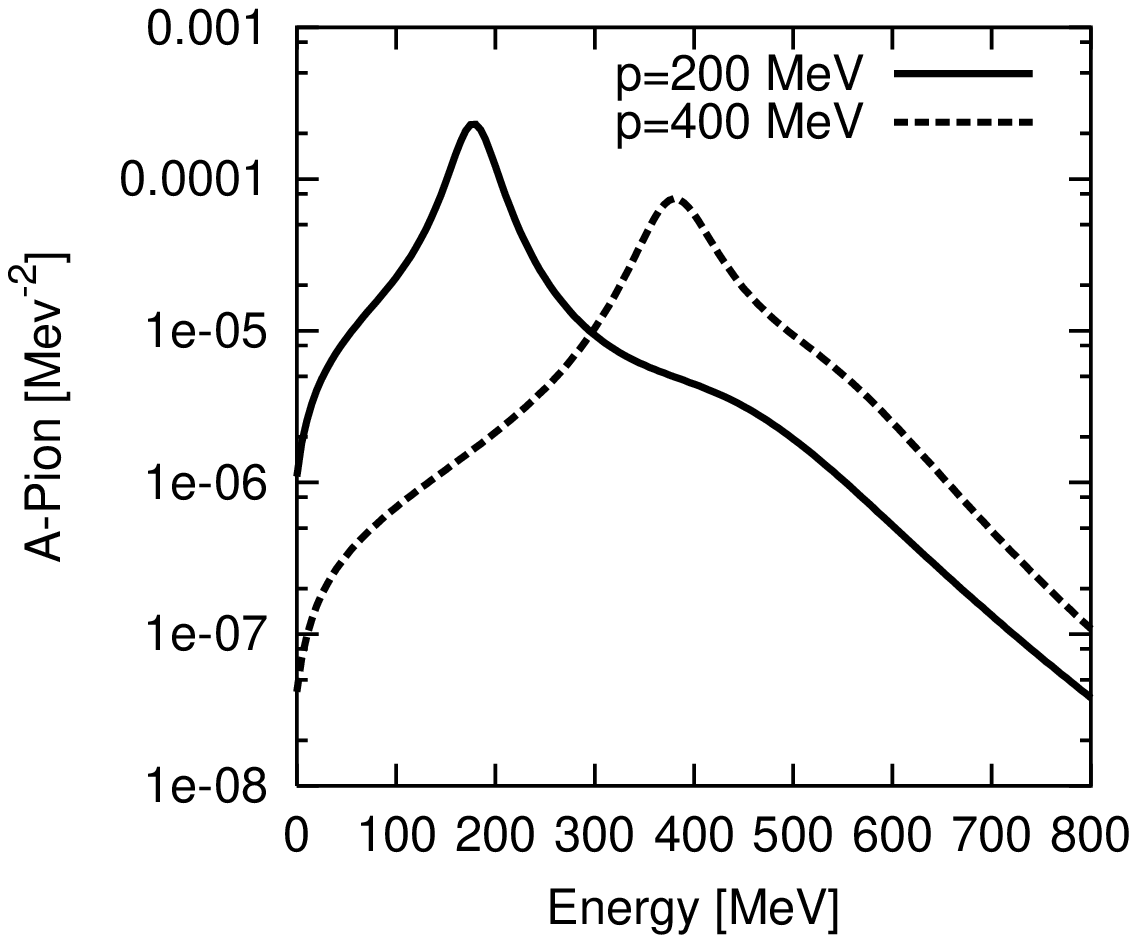,height=5 cm,width=4
      cm,angle=-90}} $ $\\
        {
{\bf Figure 3.} {Pion spectral function $ A_\pi$ at T=0
           and 120 MeV (upper/lower panel) and $\rho 
           =\rho_0$ for two different momenta.\\ $ $\label{A-pi}}}
         \end{minipage}\hfill$ $ \\
\subsection{Vector-mesons coupled to the in-medium pion cloud}
\label{Vec-mesons}

In view of the strong modifications of the pion spectral function
through the dense nuclear medium it is interesting to study its
influence on other modes that strongly couple to the pion, like the
vector mesons, i.e. the $\rho$- and $\omega$-mesons.  The vacuum
widths of these resonances are generated through the decay into two
pions in the case of the $\rho$-meson, respectively three pions for
the $\omega$-meson. For the $\omega$-meson we choose the indirect
decay via the $\rho$-meson (the so called Gell-Mann, Sharp and Wagner
(GSW) process \cite{GSW}) which is the dominant channel in vacuum.  The
coupling  Lagrangian chosen in accordance with the low energy limit
of QCD \cite{Schwinger,Wess,Witten} gives rise to the following
vector-meson polarization tensors (self-energies) 
\begin{equation}\label{SRho2Pi}\label{SOmegaRhoPi}
\unitlength0.8mm
\Pi^{\mu\nu}_{\rho}(q)=
\parbox{27\unitlength}{
\begin{fmfgraph*}(25,15)
          \fmfpen{2}
          \fmfforce{0.1w,0.5h}{i}
          \fmfforce{1w,0.5h}{o}
          \fmfforce{0.225w,0.5h}{v1}
          \fmfforce{0.875w,0.5h}{v2}
          \fmf{dbl_wiggly,fore=(0,,0.55,,0)}{i,v1}
          \fmf{dbl_wiggly,fore=(0,,0.55,,0)}{v2,o}
          \fmf{dashes,left=0.7,tension=.5,fore=red}{v1,v2}
          \fmf{dashes,left=0.7,tension=.5,fore=red}{v2,v1}
\end{fmfgraph*}}
+
\parbox{25\unitlength}{
\begin{fmfgraph*}(25,15)
          \fmfpen{2}
          \fmfforce{0.1w,0.5h}{i}
          \fmfforce{1w,0.5h}{o}
          \fmfforce{0.225w,0.5h}{v1}
          \fmfforce{0.875w,0.5h}{v2}
          \fmf{dbl_wiggly,fore=(0,,0.55,,0)}{i,v1}
          \fmf{dbl_wiggly,fore=(0,,0.55,,0)}{v2,o}
          \fmf{gluon,left=0.7,tension=.5,fore=(0,,0.55,,0)}{v1,v2}
          \fmf{dashes,left=0.7,tension=.5,fore=red}{v2,v1}
\end{fmfgraph*}}
\quad\quad
\Pi^{\mu\nu}_{\omega}(q)=
\parbox{25\unitlength}{
\begin{fmfgraph*}(25,15)
          \fmfpen{2}
          \fmfforce{0.1w,0.5h}{i}
          \fmfforce{1w,0.5h}{o}
          \fmfforce{0.225w,0.5h}{v1}
          \fmfforce{0.875w,0.5h}{v2}
          \fmf{gluon,fore=(0,,0.55,,0)}{i,v1}
          \fmf{gluon,fore=(0,,0.55,,0)}{v2,o}
          \fmf{dbl_wiggly,left=0.7,tension=.5,fore=(0,,0.55,,0)}{v1,v2}
          \fmf{dashes,left=0.7,tension=.5,fore=red}{v2,v1}
\end{fmfgraph*}}
\end{equation}
\unitlength1mm with four-momentum $q=(q_0,{\mathbf q})$. The two
vector-meson coupling vertices are adjusted to the $\pi\pi$-scattering
phase shifts together with the electromagnetic form-factor of the pion
and to the vacuum decay width of the $\omega$-meson, respectively.
For the numerical results we neglected the coupling of the
$\rho$-$\omega$-loop back to the pion as the vector mesons cause only
a minor perturbation to the $\pi N\Delta$-system.  The correlation
between $\rho$- and $\omega$-meson modes results from the
$\omega$-meson self-energy (\ref{SOmegaRhoPi}) and the reverse process
encoded in the coupling of the $\rho$-meson to the $\omega$-$\pi$-loop
which both are included self-consistently.  Since the model omits
higher lying degrees of freedom the in-medium changes of the real
parts of the vector-meson self-energies are less precisely determined
within the model space. Therefore we drop them during the
self-consistent iterations, while restoring the normalization of the
spectral functions at each step. {A posteriori, with the parameters
  used we checked for the thus ignored real parts of the loops using a
  counter-term scheme which through vector dominance fixes the photon
  mass (i.e. at $q^2=0$) to be zero with residue equal one in vacuum.
  It gave values of $\Re \Pi/(2 q_0)$ which are by far less than the
  corresponding widths of the vector mesons and therefore negligible.
  Keeping this subtraction scheme also at finite $T$, i.e. ignoring
  contributions from hidden divergences \cite{vanHeesKnoll} and
  tadpoles, which solely lead to an additional $T$-dependent mass
  shift, we still obtained insignificant changes in this quantity.}

The resulting polarization tensors (\ref{SRho2Pi})
have to be four-transversal because of current conservation. This is
provided by a strategic projection method which uses two moments of
the spacial part of the respective polarization tensors in the
self-consistent loop approximation and constructs the respective
four-transverse tensors. For details we refer to refs. \cite{Hees1,RK}.

In the following we explicitly analyze the influence of the various
components of the pion spectral function on the vector-mesons and thus
on the di-lepton spectra. Formally we do this by splitting the spectral
function of the vector-meson into the various components related to
the different processes feeding into this vector-meson channel. Thus,
decomposing the total damping width into partial widths
$\Gamma_{\rm{v},\rm{tot}}(q)=\sum_i\Gamma_{\rm{v},\rm{i}}(q)$, the
di-lepton yield can be brought into a Breit-Wigner like form with
partial in- and out-widths
\begin{eqnarray}
\frac{d R}{d^4q d^3xdt}
       &=\frac{3}{(2\pi)^4}\;f_T(q_0)\sum_{\rm{v}}A_{\rm{v}}(q)\;(-2\Im
       \Pi_{\rm{v},\rm{e}^+\rm{e^-}})\cr  
       &=\frac{3}{(2\pi)^4}\;f_T(q_0)\sum_{\rm{v},i}
       \frac {4\;q_0^2\Gamma_{\rm{v},\rm{i}}\;\Gamma_{\rm{v},\rm{e}^+\rm{e}^-}}
             {(q^2-m_{\rm{v}}^2)^2+q_0^2\Gamma_{\rm{v},\rm{tot}}^2}
\label{d R}
\end{eqnarray}

\begin{minipage}[t]{.48\linewidth}
\epsfig{file=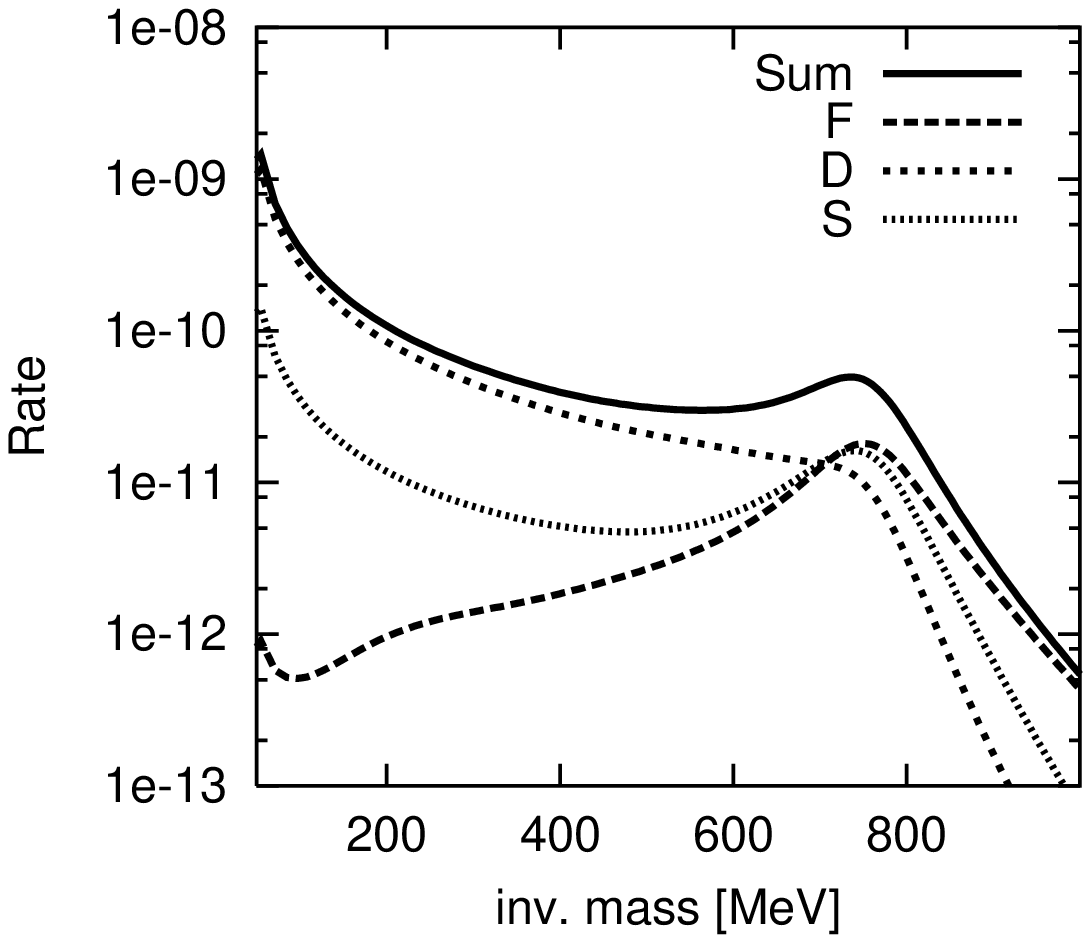,height=6.6 cm,width=6.0
  cm,angle=-90} 
\parbox{.95\linewidth}
{{\bf Figure 4.} {$e^+e^-$-rate from the decay of the $\omega$-meson at T=120
  MeV, $\rho=\rho_0$ divided into different contributions:
\label{omega-div}}}
\begin{tabular}{llll}
F:&$\pi\rho\rightarrow\omega$\\
D:&$\rho\rightarrow\pi\omega$\\
S:&$\rho N\rightarrow\omega N$.
\end{tabular}
\end{minipage}\hfill
\begin{minipage}[t]{.48\linewidth}
\epsfig{file=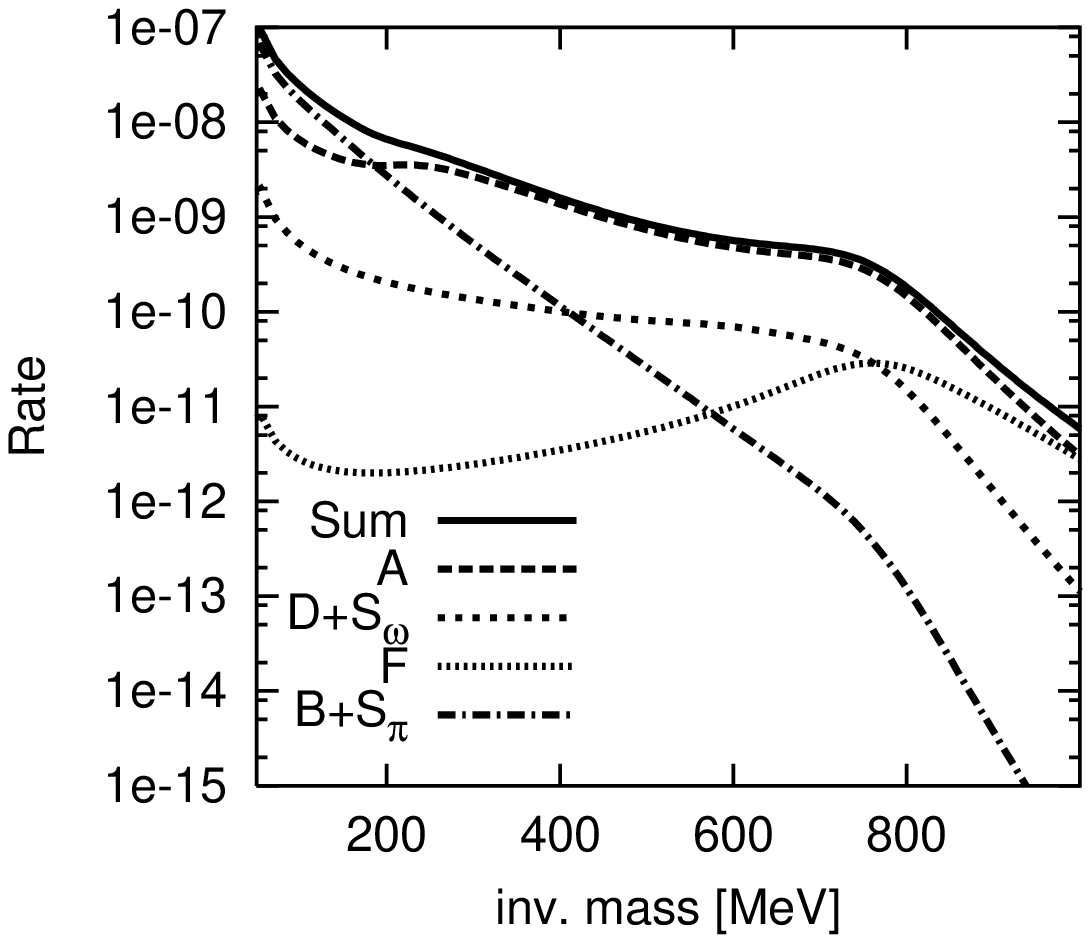,height=6.6 cm,width=6.0 cm,angle=-90}
\parbox{.95\linewidth}
{{\bf Figure 5.} {$e^+e^-$-rate from the decay of the 
$\rho$-meson at T=120 MeV, $\rho=\rho_0$ divided into  different contributions:
\label{rho-mix}}}
\begin{tabular}{llll}
A:&$\pi\pi\rightarrow\rho$&F:&$\pi\omega\rightarrow\rho$\\
B:&$\pi\rightarrow\pi\rho$&D:&$\omega\rightarrow\pi\rho$\\
S$_{\pi}$:&$\pi N\rightarrow\rho N$&S$_{\omega}$:&$\omega
N\rightarrow\rho N$.\\$ $
\end{tabular}
\end{minipage}
(suppressing the tensor structure of spectral function and
polarization tensor which leads to a degeneracy factor 3 for vector
particles).  Here $ m_{\rm{v}} $ denotes the vector meson mass,
$f_T(q_0)$ is the thermal Bose-Einstein distribution and
$\Gamma_{\rm{v},\rm{e}^+\rm{e}^-}=-\Im
\Pi_{\rm{v},\rm{e}^+\rm{e^-}}/(2q_0)$ is the di-lepton decay width of
vector-meson $\rm{v}$. In the vector dominance picture the vector
mesons directly couple to a virtual photon which then decays into the
respective lepton pairs, where for massless leptons
$\Gamma_{\rm{v},\rm{e}^+\rm{e}^-}\propto \alpha^2
m_{\rm{v}}^4/(2q_0q^2)$ with fine-structure constant $\alpha$.

We start the discussion of the effects on the $\omega$-meson. In the
medium there are three major processes contributing to its damping
width, illustrated by perturbative ``time-flow'' diagrams where the
time is running from left to right and vertical lines denote virtual
space-like propagators, which mediate two-body interactions:\\[5mm]
\setlength{\unitlength}{0.5mm}
\begin{equation}
(\rm{F})=\parbox{40\unitlength}{
\begin{fmfgraph*}(40,30)
        \fmfpen{thin}
        \fmfleft{i,i2}
        \fmfright{o}
        \fmf{dashes}{i,v1}
        \fmf{gluon}{v1,o}
        \fmf{photon}{i2,v1}
        \fmflabel{$\pi$}{i}
        \fmflabel{$\rho$}{i2}
        \fmflabel{$\omega$}{o}
\end{fmfgraph*}}
\hspace*{1cm}
(\rm{D})=\quad
\parbox{40\unitlength}{
\begin{fmfgraph*}(40,30)
        \fmfpen{thin}
        \fmfleft{i}
        \fmfright{o,o2}
        \fmf{gluon}{v1,o2}
        \fmf{dashes}{v1,o}
        \fmf{photon}{i,v1}
        \fmflabel{$\rho$}{i}
        \fmflabel{$\pi$}{o}
        \fmflabel{$\omega$}{o2}
\end{fmfgraph*}}
\hspace*{1cm}
(\rm{S})=\parbox{40\unitlength}{
\begin{fmfgraph*}(40,30)
        \fmfpen{thin}
        \fmfleft{i,i2}
        \fmfright{o,o2}
        \fmf{fermion}{i,v1}
        \fmf{fermion}{v1,o}
        \fmflabel{$N$}{i}
        \fmflabel{$N$}{o}
        \fmf{dashes,label=$\pi$,l.d=2}{v2,v1}
        \fmf{photon}{i2,v2}
        \fmf{gluon}{v2,o2}
        \fmflabel{$\rho$}{i2}
        \fmflabel{$\omega$}{o2}
\end{fmfgraph*}}
\label{om-processes}
\end{equation}\\[3mm]
\setlength{\unitlength}{1mm}%
The subsequent decay of the $\omega$-meson into the virtual time-like
photon and its final decay into the lepton pair is not illustrated. In
the self-consistent calculation with finite width spectral functions
all these processes are included automatically by one self-energy
diagram. For the first fusion-type process (F), $\rho\pi\rightarrow
\omega\rightarrow \rm{e}^+\rm{e}^-$, the $\omega$-meson is formed by
the fusion of a $\rho$-meson with a time-like in-medium pion. Its
inverse exists already in vacuum and determines the vacuum decay width
of the $\omega$-meson. The second process (D), $\rho\rightarrow
\pi\omega\rightarrow\pi \rm{e}^+\rm{e}^-$, corresponds to a
$\rho$-Dalitz-decay via an intermediate $\omega$-meson.  In the
self-consistent calculations, both above mentioned processes just
differ in the sign of the pion energy in the $\pi\rho$-loop of the
$\omega$-self-energy (\ref{SOmegaRhoPi}).  The process (S) in
(\ref{om-processes}) corresponds to the scattering
$\rho$N$\rightarrow\omega$N mediated by a virtual, i.e.  space-like
pion exchange. In view of the pion modes at zero temperature (cf. Fig.
2) we isolate this space-like component by a cut on the far space-like
region with pion loop momenta with $|\vec{p}|>2|p_0|$ in
(\ref{SOmegaRhoPi}).  At $T=120$ MeV this separation is less evident
in view of the broad structure of the pion spectral function (Fig. 3).
Thus the different components of the processes displayed in Fig. 4
somewhat depend on this cut. In addition to the fusion width (F),
which constitutes just a temperature dependent modification of the
vacuum width (e.g. accounted for by Schneider and Weise
\cite{Schneider1}) a genuine in-medium process, namely the scattering
process (S), contributes with comparable strength at the nominal
resonance position. The ``Daliz''-decay of the $\rho$-meson (D)
dominates the low mass region.

In summary of this analysis: a major portion to the $\omega$
spectrum results from processes (S) which are not accounted for in the
simple on-shell treatment. These contributions, however, sensitively
depend on the in-medium properties of the virtual pion cloud, which
certainly needs further clarifying investigations before quantitative
conclusions can be drawn.

The same type of processes as in (\ref{om-processes}) also occur for
the $\rho$-meson just interchanging $\rho$ with $\omega$, listed in
the $\rho$-meson decomposition given in Fig. 5 as (F), (D)
and (S$_{\omega}$). However, these $\omega$-induced components are less
important compared to the coupling to the two-pion channels given by

\setlength{\unitlength}{0.5mm}
\begin{equation}
(\rm{A})=\parbox{40\unitlength}{
\begin{fmfgraph*}(40,30)
        \fmfpen{thin}
        \fmfleft{i,i2}
        \fmfright{o}
        \fmf{dashes}{i,v1}
        \fmf{photon}{v1,o}
        \fmf{dashes}{i2,v1}
        \fmflabel{$\pi$}{i}
        \fmflabel{$\pi$}{i2}
        \fmflabel{$\rho$}{o}
\end{fmfgraph*}}
\hspace*{1cm}
(\rm{B})=\quad
\parbox{40\unitlength}{
\begin{fmfgraph*}(40,30)
        \fmfpen{thin}
        \fmfleft{i}
        \fmfright{o,o2}
        \fmf{photon}{v1,o2}
        \fmf{dashes}{v1,o}
        \fmf{dashes}{i,v1}
        \fmflabel{$\pi$}{i}
        \fmflabel{$\pi$}{o}
        \fmflabel{$\rho$}{o2}
\end{fmfgraph*}}
\hspace*{1cm}
(\rm{S}_{\pi})=\parbox{40\unitlength}{
\begin{fmfgraph*}(40,30)
        \fmfpen{thin}
        \fmfleft{i,i2}
        \fmfright{o,o2}
        \fmf{fermion}{i,v1}
        \fmf{fermion}{v1,o}
        \fmflabel{$N$}{i}
        \fmflabel{$N$}{o}
        \fmf{dashes,label=$\pi$,l.d=2}{v2,v1}
        \fmf{dashes}{i2,v2}
        \fmf{photon}{v2,o2}
        \fmflabel{$\pi$}{i2}
        \fmflabel{$\rho$}{o2}
\end{fmfgraph*}}
\label{rho-processes}
\end{equation}\\
\setlength{\unitlength}{1mm}%

 At invariant masses above 300 MeV, the
$\pi^+\pi^-$-annihilation process (A) is clearly dominant. Processes
(B) and ($S_\pi$) are not present in any on-shell treatment of the pion as
they solely arise from genuine off-shell components of the pion
spectral function.  In this respect (B) can be interpreted as a
bremsstrahlung process radiated off a pion scattered in the medium,
while process (S) corresponds to inelastic $\pi N\rightarrow \rho N$
scatterings mediated via virtual pion exchange. The latter two
components, which emerge completely consistently within the model,
only contribute to the very low-mass part of the invariant mass
spectrum.
\section{Conserving non-equilibrium dynamics of particles and
  resonances}
The equilibrium results discussed above can immediately be implemented
into a hydrodynamical scheme which relies on the assumption of local
equilibrium. This concerns the EoS derived from the corresponding
thermodynamic potential, as well as the spectral properties of the
particles and resonances as the electromagnetic decay rates. As
penetrating probes the latter can simply be integrated over the
hydrodynamic space-time evolution, while strongly interacting probes
require an appropriate freeze-out concept.

On the microscopic level the Kadanoff-Baym (KB) equations \cite{KB}
provide the appropriate framework for a self-consistent dynamical
treatment.  However, only very recently first attempts to solve these
equations for very simple scalar models were successfully under taken
\cite{JCG,Berges} in admittedly very simplified cases of simple scalar
fields in spatially homogeneous systems. For the complex situations in
nuclear dynamics substantial progress in computing power is required
before meaningful calculation with the KB-equations can be performed.
Therefore most application are dealing with transport approaches where
a Boltzmann-like transport equation is solved for the involved
microscopic degrees of freedom. The numerical solution is mostly done
by Monte Carlo simulation methods.

In this context one of the key issues was to generalize the concepts
such that they can also properly deal with resonances or more generally
with particles which have non-trivial i.e. non quasi-particle like
spectral functions. The corresponding concepts were already laid out
in the book of Kadanoff and Baym \cite{KB}. It involves a systematic
gradient approximation of the Wigner transformed KB-equations,
avoiding the step towards the quasi-particle approximation.  Thereby
the question of thermodynamic consistence and conservation laws is of
vital importance. Is such a transport scheme thermodynamically
consistent and are the conservation laws associated with the
symmetries of the underlying theory fulfilled?

\subsection{$\Phi$-derivable scheme and exact conservation laws}
\label{Phi-Conserv} 

For self-consistent Dyson-resummation schemes and thus for the
KB-equations the question of thermodynamic consistence and
conservation laws was settled by Baym \cite{Baym}. He showed that such
schemes are conserving, if and only if all self-energies result from
the functional variation of a functional, called $\Phi$, which is a
functional of the self-consistent classical fields (one-point
functions) and two-point propagators, see also \cite{IKV99}. The
functional method initially introduced by Luttinger and Ward \cite{LW}
and later reformulated in terms of path integrals became known as the
CJT-formalism \cite{CJT} in field theory according to the names of the
authors. The $\Phi$-functional is given by two-particle irreducible
(2PI) closed diagrams in accordance with the Lagrangian of the system
expressed in terms of the self-consistent propagators and classical
fields, while the vertices remain the bare ones.

With focus on the Kadanof-Baym equations we discard the dynamics
of the classical field for the further considerations and concentrate
on the self-consistent dynamics of the propagators $G(x,y)$.  In a
$\Phi$-derivable scheme the self-energies are generated from the
functional $\Phi\{\Gr,\lambda\}$ through the following functional
variation\footnotemark, cf.  \cite{Baym,IKV99}
\begin{eqnarray}\label{varphdl1}
-\ii \Se(x,y)
&=&\mp\frac{\delta\ii \Phi\{\Gr,\lambda\}}{\delta \ii\Gr(y,x)}\times
\left\{
\begin{array}{ll}
2\quad&\mbox{for real fields}\\
1\quad&\mbox{for complex fields}
\end{array}\right. \quad \mbox{where}\\
\ii\Phi\{\Gr,\lambda\}&=&
\left<\exp\left(\ii\oint\di^4 x \lambda(x) {\cal L}^{\rm
      int}\right)\right>_{2PI} 
\label{Phi-def}
\end{eqnarray}
%
(upper/lower signs refer to fermions/bosons).%
Here and in the following labels for the different species in the
system and internal quantum numbers are suppressed.  Taking all 2PI
diagrams for $\Phi$ provides the exact self-energy and two-point
propagator of the theory.  The space dependent coupling factor
$\lambda(x)$, whose physical value is unity, is introduced to permit
further variations, cf. Eq.  (\ref{eps-int}) below.

The virtue of the $\Phi$-derivable scheme (\ref{varphdl1}) is, that
the diagrammatic series (\ref{Phi-def}) of $\Phi$ can be truncated at
any level, without spoiling the conserving properties of the
self-consistent scheme (\ref{varphdl1}) for the Dyson-Schwinger or KB
equations, see Eq. (\ref{KB-Eq}) below.  The so constructed
self-energies lead to a coupling between the different species in
the system, which obey detailed balance generalizing the special
recipes for broad resonances given in ref.\cite{DB91}. It has been
shown \cite{Baym}, see also \cite{IKV99}, that such a self-consistent
scheme is thermodynamically consistent and exactly conserving at the
expectation value level of the conserved currents.

For local couplings the interaction energy density is given by a
functional variation with respect to the interaction strength
$\lambda(x)$
\begin{eqnarray}
\label{eps-int} 
{\cal E}^{\scr{int}}(x)&=&\left<-{\cal L}^{\rm
      int}(x)\right>
=-\left.\frac{\delta\ii\Phi}{\delta\ii\lambda(x)}\right|_{\lambda=1}.
\end{eqnarray}
The single-particle potential energy density is defined as
\begin{eqnarray}
\label{eps-pot}
{\cal E}^{\scr{pot}}(x)
&=& \frac{1}{2}
\oint\di^4 y \left[
\Se(x,y) (\MP\ii)\Gr(y,x)+(\MP\ii)\Gr(x,y)\Se(y,x)\right]  
\end{eqnarray}
here written for complex fields. Both, ${\cal E}^{\scr{int}}$
and 
${\cal E}^{\scr{pot}}$, enter the energy--momentum tensor (\ref{E-M}).

The $\Phi$-derivable scheme (\ref{varphdl1}) implies exact
conservation laws for the Noether currents and the energy--momentum
tensor \cite{IKV99} given by
\begin{eqnarray}
\label{Q-E-M}
\partial_{\mu}J^{\mu}(X)&=&0,\quad\partial_{\mu}
\Theta^{\mu\nu}(X)=0
\hspace*{1cm}\mbox{with}
\\
J^{\mu}(X)&=&\sum_a\intp{e}v^{\mu}(\mp\ii)F(X,p),\label{Q-Noether}\\
\Theta^{\mu\nu}(X)
&=&\intp v^{\mu}p^{\nu}(\mp\ii)F(X,p) 
+g^{\mu\nu}\left(\E^{\mathrm{int}}(X)-
\E^{\mathrm{pot}}(X)\right).\label{E-M}
\end{eqnarray}
Here 
\begin{equation}
F(X,p)=\mp\ii G^{-+}(X,p)\quad  (\mbox{alias}\quad \mp\ii G^{<}(X,p))
\end{equation}
denotes the distribution function in four-phase-space given by the four-Wigner
transformation of the two-point propagators $G^{-+}(x_1,x_2) $, while
$e$ stands for any of the conserved charges and the kinematical factor
$ v^{\mu}$related to the four-velocity is define in (\ref{vmu}) below.

\subsection{Kadanoff--Baym equations and complete gradient approximation}
\label{QKE}
We assume the reader to be familiar with the real-time formulation of
non-equilibrium field theory on the so called closed time contour,
ubiquitously used in this book.  Since we will deal with general multi-point
functions we use the more convenient ``$-+$'' contour-vertex notation
of refs.  \cite{LP,IKV99} rather than the more clumsy ``$<$ $>$''
notation.  In the following we follow the lines given in ref.
\cite{KIV-Annals}.  The set of coupled KB equations on the time
contour in ``$-+$'' notation\footnote{The numbers 1, 2 and 3 provide
  short hand notation for space--time coordinates $x_1,x_2$ and $x_3$,
  respectively, including internal quantum numbers. With superscript,
  like $1^-$ and $2^+$, assigned to them, they denote contour
  coordinates with $-$ and $+$ specifying the placement on the time or
  anti-time ordered branch. For an arbitrary two-point function $f$
  its decompositions into the two branches of the contour are denoted
  as $f^{kl}(1,2)=f(1^k,2^l)$ with $k,l\in \{-,+\}$. The match to the
  notation used, e.g., in refs.  \cite{KB,D} is given by
  $f^{-+}=f^{<};\; f^{+-}=f^{>};\; f^{--}=f^c;\; f^{++}=f^a$.}  reads
\begin{eqnarray}\label{KB-Eq}
\left(G^{-1}_{0}(-\ii\partial_1)-G^{-1}_{0}(-\ii\partial_2)\right)
\Gr^{-+}(1,2) 
&=&\oint\di 3 \left(\Se(1^-,3 )\Gr(3 ,2^+) -\Gr(1^-,3 )\Se(3,2^+)\right)\cr 
&\equiv& \MP C(1^-,2^+)
\end{eqnarray}
and likewise for $\Gr^{+-}$. Here 
\begin{eqnarray}
\Gr^{-1}_{0}(p)=\left\{
\begin{array}{ll}
p^2-m^2\quad&\mbox{for relativistic bosons}\\
p_0-{\vec p}^2/(2m)\quad&\mbox{for non-rel. fermions or bosons.}
\end{array}\right.
\end{eqnarray}
is the inverse free Green function in momentum representation. The
upper/lower sign factors refer to fermions or bosons, respectively,
$\Gr_0$ and $\Gr$ correspondingly denote the free and full Green
functions. The driving term on the r.h.s. of Eq.
(\ref{KB-Eq}), summarized by $C$, is a functional of the Green
functions through the self-energies $\Se$ contour folded with $\Gr$.
The real-time integration contour is denoted by ${\cal C}$.  

As usual the step towards transport equations is provided by
introducing the four-dimensional Wigner transforms for any two-point
function $f$ through
\begin{eqnarray}
\label{Wigner}
f(x,y)=\intp e^{-\ii p(x-y)} f({\textstyle\frac{x+y}2},p).
\end{eqnarray}
or its inverse relation.
The KB Eq.  (\ref{KB-Eq}) then transforms to
\begin{eqnarray}
\label{KB-Eq-Wigner} 
v^{\mu}\partial_{\mu}F(X,p)&=& C^{-+}(X,p;\{\Gr\})\\
\quad\mbox{with}\quad
F(X,p)&=&\MP\ii \Gr^{-+}(X,p)
\quad\mbox{and}\quad\label{vmu}
v^{\mu}=\frac{\partial}{\partial p_{\mu}}G_0^{-1}(p).
\end{eqnarray}
Here the r.h.s. of the KB equation is also expressed in terms of the
Wigner transforms of all Green functions through (\ref{Wigner}).  The
final step is to expand the complicated r.h.s. of Eq.
(\ref{KB-Eq-Wigner}) to the first-order in space-time gradients, cf.
\ref{Gadient-Apprx} or \cite{KIV-Annals} for more details. Then the
{\em local} part of this r.h.s., $C^{-+}_{\mathrm{(loc)}}$, defines
the collision term. It entirely consists of {\em non-gradient terms},
where all the different mean positions $(x_i +x_j )/2$ occurring in
the various Wigner-Green functions are replaced by the externally
given mean position $X$ of the l.h.s., i.e.  $X=(x_1+x_2)/2$ and the
diagrams are to be evaluated as in momentum representation. The
corrections for the displacement to the true coordinates of each Green
function are then accounted for to first order in space-time
gradients.  Here we simply abbreviate the gradient terms by a
$\Diamond$ operator defined in \ref{Gadient-Apprx} acting on the {\em
  local} diagram expression
\begin{eqnarray}
\label{Grad-KB-Eq}
v^{\mu}\partial_{\mu}(\MP\ii)\Gr^{-+}(X,p)
&=&  (1+{\textstyle\frac{\ii}{2}}\Diamond)
\left\{C^{-+}_{\mathrm{(loc)}}(X,p)\right\},
\end{eqnarray}
where
\begin{eqnarray}
\label{C-loc}
C^{-+}_{\mathrm{(loc)}}(X,p)&=&C^{-+}(X,p;\{\Gr_{\mathrm{(loc)}}\})
,
\end{eqnarray}
is a functional of the local Green functions $\Gr_{\mathrm{(loc)}}
\equiv \Gr(X,p)$.  Here and for all further considerations below,
both, Green functions $\Gr(X,p)$ and self-energies $\Se(X,p)$,
whenever quoted in their Wigner function form, are taken in {\em
  local} approximation, i.e. with $X$ given by the external coordinate
and $\Se(X,p)$ void of any gradient correction terms. The gradient
terms (seem \ref{Gadient-Apprx}) have the form that in the diagrams
defining $C_{\mathrm{(loc)}}$ pairs of Green functions are replaced by
their space-time and momentum derivatives, respectively, just leading
to equations linear in space-time gradients. Naturally the result of
the $\Diamond$-operation depends on the explicit form, i.e.
diagrammatic structure, of the functional on which it operates.
For the non-gradient term in Eq. (\ref{Grad-KB-Eq}), which
defines the {\em local} collision term
\begin{eqnarray}
\label{C-loc-dia}
C^{-+}_{\mathrm{(loc)}}
&{\begin{array}[t]{c}=\\[-1mm]
{\!\!\!\!\!_\scr{diagram}\!\!\!\!\!}\end{array}}
&
\MP \Se^{-k}(X,p)\sigma_{kl}\Gr^{l+}(X,p)
-(\MP) \Gr^{-k}(X,p)\sigma_{kl}\Se^{l+}(X,p)\\
\label{C-loc-value}
&{\begin{array}[t]{c}=\\[-1mm]
{\!\!\!_\scr{value}\!\!\!}\end{array}}
&
\,\underbrace{\MP \ii\Se^{-+}(X,p)\ii\Gr^{+-}(X,p)}_{\mbox{gain}}
\, -\,
\underbrace{(\MP)\ii\Gr^{-+}(X,p)\ii\Se^{+-}(X,p)}_{\mbox{loss}}
,
\end{eqnarray}
we give
both, the diagram expression (\ref{C-loc-dia}) and the normally quoted {\em
value} expression (\ref{C-loc-value}). The latter simplifies due to a
cancellation of terms which however survive for the order sensitive gradient
operation. 
Here $\sigma_{ik}=\sigma^{ik}=\mathrm{diag}(1,-1)$ defines the ``contour
metric'', which accounts for the integration sense, and summation over the
contour labels $k,l\in \{-,+\}$ is implied.

The above quantum kinetic equation (\ref{Grad-KB-Eq}) has to be supplemented
by a {\em local} Dyson equation for the retarded Green function \cite{KB}
\begin{eqnarray}\label{retarded-Eq}
\left(\Gr^R(X,p)\right)^{-1}=\left(\Gr^R_0(p)\right)^{-1}-\Sigma^R(X,p),
\end{eqnarray}
which together with Eq. (\ref{Grad-KB-Eq}) provides the simultaneous
solution to $\Gr^{+-}$. The full retarded Green function $\Gr^R$
depends on the retarded self-energy
$\Se^R=\Se^{--}-\Se^{-+}=\Se^{+-}-\Se^{++}$ again in {\em local}
appro\-ximation.  $\Gr^R_0$ is the free retarded Green function.
Please note that equation (\ref{retarded-Eq}) is just algebraic
although it is obtained within the frame of the first-order gradient
approximation.

In most presentations of the gradient approximation to the KB
equations, Eq.  (\ref{Grad-KB-Eq}) is rewritten such that the gradient
terms are subdivided into different physical parts: on the one hand
Poisson bracket terms describing drag- and back-flow effects which are
pulled to the left side of the equation and in cases where the
self-energy explicitly contains internal vertices an additional
memory collision term $C^{\scr{mem}}$
\begin{eqnarray}\label{Grad-Sep-KB-Eq}
v^{\mu}\partial_{\mu}F(X,p)+
\Pbr{\Re\Se^R,\pm\ii\Gr^{-+}}&+&\Pbr{\pm\ii\Se^{-+},\Re\Gr^{R}}
=C^{-+}_{\mathrm{(loc)}}(X,p)+ C^{-+}_{\mathrm{(mem)}}(X,p).\hspace*{5mm}\\
\mbox{with}\quad\label{C-mem}\nonumber
\MP C^{-+}_{\mathrm{(mem)}}(X,p)
&=&\Se^{-k}_{(\scr{mem})}(X,p)\sigma_{kl}\Gr^{l+}(X,p)
-\Gr^{-k}(X,p)\sigma_{kl}
\Se_{(\scr{mem})}^{l+}(X,p)
\\
\label{C-mem-value}
&\!\!\begin{array}[t]{c}=\\[-1mm]
{\scr{value}}\end{array}\!\!&
{-\Se_{(\scr{mem})}^{-+}(X,p)\Gr^{+-}(X,p)}
+
{\Gr^{-+}(X,p)\Se_{(\scr{mem})}^{+-}(X,p)},
\end{eqnarray}
where $\Se_{(\scr{mem})}(X,p)=
{\textstyle\frac{\ii}{2}}\Diamond\left\{\Se(X,p)\right\}$
is only non-zero if internal vertices exist in $\Se$.

Depending on the questions raised either the here presented separation
of the gradient terms or the compact formulation (\ref{Grad-KB-Eq}) is
more advantageous. For the derivation of the conservation laws only a
unified treatment of both, the Poisson brackets and memory collision
terms, reveals the symmetry among these terms, which then displays the
necessary cancellation of certain contributions such that the
conservation laws emerge. Starting form Eq. (\ref{Grad-KB-Eq}) and
using the diagrammatic gradient rules given in \ref{Gadient-Apprx} one
could indeed prove that a) the Noether currents (\ref{Q-Noether}) and
b) the {\em local} approximation of the energy-momentum tensor
(\ref{E-M}) are exactly conserved. In particular for the latter
derivation the diagram rules (\ref{x-dashed}) to (\ref{Poisson}) where
an essential help to isolate the proper counting of diagrams with
terms proportional to the number of vertices contributing to
$\E_\scr{int}$ in (\ref{E-M}) and those proportional to the number of
Green functions contributing to $\E_\scr{pot}$. 

\subsection{Physical interpretation of terms}

For the physical interpretation it is advantageous to introduce the
spectral function $A(X,p)=-2\Im \Gr^R(X,p)$ determined by the retarded
equation (\ref{retarded-Eq}), as well as the four-phase-space
occupation functions $f(X,p)$, by means of $F(X,p)=f(X,p) A(X,p)$. In
thermal equilibrium $f(X,p)$ becomes a Fermi-Dirac or Bose-Einstein
distribution in the particles energy $p_0$. The evolution of $F$ is
governed by transport Eq.  (\ref{Grad-Sep-KB-Eq}) or equivalently by
(\ref{Grad-KB-Eq}). Together with the retarded equation
(\ref{retarded-Eq}) this defines a generalized quantum transport
scheme which is void of the usual quasi-particle assumption.  The
space-time evolution is completely determined by the initial values of
the Green functions at time zero for each space point. Thus the
evolution is ``{\em Markovian}'', since the memory part of the
collision term is kept up to first-order gradient terms only. Within
its validity range this transport scheme is capable to describe slow
space-time evolutions of particles with broad damping width, such as
resonances, within a transport dynamics, now necessarily formulated in
the four-dimensional phase-space.

Coming from the usual on-shell Boltzmann or
Boltzmann-Uehling-Uhlenbeck collision term, each occurring
three-momentum distribution function together with its momentum
integration is simply to be replaced by its four-momentum analog,
thus
\begin{equation}
f(X,{\vec{p}}) \frac{\di ^3{\vec{p}}}{(2\pi)^3}
\Longrightarrow
F(X,p) \frac{\di ^4 p}{(2\pi)^4}=
f(X,p) A(X,p) \frac{\di ^4 p}{(2\pi)^4}.
\end{equation}
Alongside the normally occurring two-body cross-sections have to be
replaced by the corresponding $T$-matrix expressions providing the
proper off-shell extensions. For genuine momentum dependent
$T$-matrices the collision term has a finite virial due to the
interactions at finite distances and therefore the collision term
contributes to the conservation laws in a non-trivial fashion. 
Within field theory model applications with local couplings
one simply has to evaluate the corresponding self-energies for the
collision term. In this case the local collision term drops out of the
conservation laws. With $\Phi$-derivable self-energies the collision
term assures detailed balance.

More subtle are the first order gradient terms given by the two
Poisson brackets and the memory collision term in Eq.
(\ref{Grad-Sep-KB-Eq}). All three terms contribute to the conservation
laws. Thereby the first Poisson bracket furnishes the so-called
drag-flow. In the quasi-particle limit it accounts for the dressing of
the particles by the dragged matter cloud as to form a quasi-particle
with a non-trivial dispersion relation with a corresponding in-medium
group velocity that can be expressed by an effective mass. This change
in flow is just compensated by the second Poisson bracket through the
polarization of the medium. The latter thus forms a back-flow
component. Only the coherent play of both Poisson brackets restores
the conserved Noether currents and thus recovers e.g.  Galilei (Lorentz)
invariance.

Since the first Poisson bracket involves space-time and momentum
gradients directly acting on the distribution function $F$ this term
has an easy classical interpretation where the motion of the
corresponding particle is subjected to a force which generally is
momentum dependent. A generalization of this concept to the
four-momentum picture is straight forward, since it just amounts to
establish the corresponding characteristic curves of the homogeneous
first-order differential equation. For the second Poisson bracket term
on the other hand the derivatives of the distribution function appear
only implicitly through the self-energy with the result that they
affect momenta other than the momentum externally entering the
transport equation. This has to be such, since the discussed term
describes the reaction of the surrounding matter on the particle
moving through the matter.  However this term escapes an immediate
description in terms of test particles, such that a simulation
algorithm could not yet be established for the exact quantum kinetic
equation (\ref{Grad-Sep-KB-Eq}).

Guided by equilibrium relations Botermanns and Malfliet \cite {BM}
suggested a simplification of this second Poisson term, cf.
\cite{IKV00,Leupold,CJ}
\begin{equation}
\Pbr{-\ii\Se^{-+},\Re\Gr^{R}}
{\begin{array}[t]{c}\Longrightarrow\\[-1mm]
{\!\!\!\!\!_\scr{BM}\!\!\!\!\!}\end{array}}
\Pbr{f(X,p) \Gamma(X,p),\Re\Gr^{R}}\quad\quad\mbox{with} \quad
\Gamma(X,p)=-2\Im \Se^{R},
\end{equation}
formally valid up to second order gradient terms. Here the
distribution function $f(X,p)$ directly appears, while $\Gamma$ is the
damping width. The advantage of this substitution is that now the
Poisson-bracket derivatives directly act on the distribution function
$f$ and the term amends a test-particle simulation\cite{Leupold,CJ}.
The price to be payed is that then the conservation laws are slightly
modified, since instead of the spectral function $A$ rather the
entropy-spectral function $A_s=\frac{1}{2} A\Gamma^2$, as introduced
in ref.  \cite{Carneiro}, enters the conserved current expression
\begin{equation}
J_{\rm BM}^{\mu}(X)=\intp{e}v^{\mu}f(X,p)A_s(X,p)(X,p).\label{Q-BM}
\end{equation}
The BM-substitution accounts for part of the back-flow. In the
quasi-particle limit both spectral functions converge to the same
$\delta$-function at the quasi-particle energy \cite{Carneiro,IKV00}.
A further merit of the BM-substitution is that for certain collision
terms an entropy current can be derived which fulfills an exact
H-theorem, for details see ref. \cite{IKV00}.

\section{Summary and prospects}
The initially presented nuclear physics examples show that in the
dense nuclear environment the particles acquire strongly broadened
spectral functions. They reach damping widths which are of the same
order as the systems temperature or even larger. Such spectral functions
escape the conventional treatment within the quasi-particle limit. 

For the application to nuclear-collisions dynamics problems the
Kadanoff-Baym equations within a certain truncation level would
certainly be the ultimate description level. However the required
numerical afford presently limits such application to simple
space-homogeneous systems and prevents realization in full three space
dimension, time and four momentum space! Therefore we review the
prospects of generalized transport equations derived in first-order
gradient approximation from the KB-equations. Avoiding the
quasi-particle limit the resulting set of dynamical equations consists
of a generalized transport equation and a retarded equation for each
type of particle in the system.

We conclude that the resulting equations posses remarkable properties
if derived in a so called $\Phi$-derivable scheme.  There the driving
terms, the self-energies, are generated from a functional of closed
diagram, the $\Phi$-functional, of the non-equilibrium Green functions
that can be truncated at any level.  Thereby the transport equation
consists of a local collision term which obeys detailed balance and
first order gradient terms.  In their exact form the first order
gradient terms guarantee exactly conserved Noether currents and a
conserved energy-momentum tensor (the latter provided that local
couplings are used). In thermal equilibrium they furnish a non-trivial
equation of state in accordance with the truncation level of $\Phi$.
The equations of motion further preserve the retarded properties of
the real-time components of the contour Green functions.  Thus, the
dynamical equations provide a {\em generic} scheme for a proper
treatment of the motion of particles and resonances with non-trivial,
i.e. finite mass-width spectral functions. Certain gradient terms
escape an immediate numerical simulation in terms of classical test
particles.  However first numerical applications \cite{JCG} with the
Botermanns-Malfliet substitution \cite{BM,IKV00,CJ} look
promising.

Initially we aimed at a proper transport description for broad
resonances.  However the presented schemes with dynamical spectral
functions offers even more. Already at the lowest order self-energy
level (genuine two-point function with no further internal vertices),
which leads to a collision term and therefore to a finite damping
width, the scheme provides the following advantages:
\begin{itemize}
\item the scheme includes higher order processes such as
  Bremsstrahlung which would be absent at the very same diagram level
  in the quasi-particle approximation;
\item it leads to completely regular expressions at any order of the
  self-energy; in order to avoid double counting and the inclusion of
  redundant processes which are generated from lower order processes
  through the iteration of the equations of motion, higher order
  self-energy terms have just simply to be derived from a 2PI
  generating functional;
\item therefore it avoids serious conceptual problems that arise in
  the use of the quasi-particle approximation for higher order
  processes (multi-particle scattering). The latter lead to
  mathematically ill defined expressions (squares of delta-functions
  or principle values) for the collision term which precisely arise
  whenever an intermediate propagator kinematically acquires on-shell
  conditions.  The latter processes are precisely the reducible ones,
  as they can be generated through an iteration of lower order
  processes. Within the quasi-particle picture there is no clear cut
  way to avoid such complications and mathematical pathologies. Only
  partial resummation schemes such as the Dyson resummation with
  finite damping width spectral functions can cure this problem.
\end{itemize}
The partial resummations implied by the self-consistent Dyson or KB
scheme lead to violations of Ward-Takahashi identities for higher
order correlation functions. In the context of vector mesons this
affects the current conservation of the polarization tensor. As we
briefly discussed, special projection strategies \cite{Hees1,RK} can
be used to recover the four transversality. Also in cases of
spontaneously broken symmetries the Nambu-Goldstone theorem is
violated as the self-consistent Goldstone-boson masses are non zero.
Recently the authors \cite{IRK05} suggested a correction term to the
$\Phi$-functional which recovers massless Goldstone bosons in the
broken phase.

In conclusion, the generalized transport equations together with the
$\Phi$-derivable concept represent a generic dynamical scheme to deal
with particles which attain broad spectral functions.  In addition to
the merits of this approach there are prospects that some of the
deficiencies can be cured by adapted corrections without leaving this
conceptual frame. E.g. with applications to Quantum Chromodynamics
(QCD) in mind this gives hope that at a proper time also non-trivial
self-energies can be treated for the case of gauge fields.

\section*{Acknowledgments}
We are grateful to G. Baym, J. Berges, P. Danielewicz, B. Friman, H.
van Hees, C. Greiner, E.E. Kolomeitsev, M. Lutz and S. Leupold for
fruitful discussions on various aspects of this research. Two of us
(Y.B.I. and D.N.V.) highly appreciate the hospitality and support
rendered to us at Gesellschaft f\"ur Schwerionenforschung.  This work
was supported in part by the Deutsche Forschungsgemeinschaft (DFG
project 436 RUS 113/558/0-2) and the Russian Foundation for Basic
Research (RFBR grant 03-02-04008).

\appendix

\section{Diagram rules for the gradient approximation}
\label{Gadient-Apprx}

Let $M(1,2)$ be any two-point function with complicated internal
structure. We are looking for its Wigner function $M(X,p)$ with
$X=\frac{1}{2}(x_1+x_2)$ to first-order gradient approximation. The
zero-order term is just given by evaluating $M(1,2)$ with the Wigner
functions of {\em all} Green functions taken at the same space-time
point $X=(x_1+x_2)/2$ and the momentum integrations being done as in
the momentum representation of a homogeneous system. To access the
gradient terms related to any Green function $G(i,j)$ involved in
$M(1,2)$, its Wigner function $G(\frac{1}{2}(x_i+x_j),p)$ is to be
Taylor expanded with respect to the space-time reference point
$X=(x_1+x_2)/2$, i.e.
\begin{eqnarray}
G(\frac{x_i+x_j}{2},p)\approx G(X,p)
+\frac{1}{2}\left[(x^{\mu}_i-x_1^{\mu})+(x_j^{\mu}-x_2^{\mu})\right]
\frac{\partial}{\partial X^{\mu}}G(X,p). 
\end{eqnarray}
Both,  $\partial_X G(X,p)$ and the accompanying factors $(x_i-x_1)$
and $(x_j-x_2)$, can be taken as special two-point functions,
to which special diagrams can be assigned\unitlength0.75mm
\begin{eqnarray}\label{x-dashed}
\Gdashed
\;&=&\;{\textstyle\frac{1}{2}}\left(\partial _i +\partial_j\right)\Gr(i,j)
\longrightarrow \partial_X \Gr(X,p), \\
\label{p-dashed}
\Pdashed{i}{j}
\;&=&\;-\ii\left(x_i-x_j\right)
\longrightarrow -(2\pi)^4\frac{\partial}{\partial p}\delta(p) 
\end{eqnarray}
with the corresponding Wigner functions on the right hand side. Then
the gradient terms of any complicated two-point function (given in
different notations) can graphically be represented by the following
two diagrams on the r.h.s.
\begin{eqnarray}\label{Gradient-Diag}
\Diamond \left\{M(1,2)\right\}&=&
\Diamond
\FBox{$1$}{$2$}{$M$}\!
\equiv\!\FBox{$1$}{$2$}{$\Diamond M$}
\!=\!
\FBoxL+\FBoxR
\end{eqnarray}
Here the diamond operator $\Diamond$ formally defines the spatial
gradient approximation of the two-point function $M$ to its right with
respect to the two external points $(1,2)$ displayed by full dots. The
diagrammatic rules are then the following. For any $\Gr(3,4)$ in $M$,
take the spatial derivative $\partial_X\Gr(X,p)$ (double line) and
construct the two diagrams, where external point 1 is linked to 3 by
an oriented dashed line, and where point 2 is linked to 4,
respectively. Interchange of these links provides the same result.
Here $M'$ is a four-point function generated by opening $M(1,2)$ with
respect to any propagator $G(3,4)$, i.e.
\begin{eqnarray}
M'(1,2;3,4)=\MP\frac{\delta M(1,2)}{\delta \;\ii G(4,3)}.
\end{eqnarray}
The diagrams in Eq. (\ref{Gradient-Diag}) are to be evaluated in the
local approximation, i.e. with all Wigner Green functions taken at the
externally given space-time point $X$. The dashed line
(\ref{p-dashed}) adds a new loop to the diagram, which, if integrated,
leads to momentum derivatives of the Green functions involved in that
loop. Both, double and dashed lines have four-vector properties, and
the rule implies four-scalar products between them.  From
(\ref{p-dashed}) follows that \unitlength0.75mm
\begin{eqnarray}\label{p-dashed-addition}
\Pdashed{1}{3}\;&=&\;\Pdashed{1}{2}\;+\;\;\Pdashed{2}{3}
\quad\quad{\rm and}\quad\quad
\Ploop
=0.
\end{eqnarray}
In momentum representation these rules correspond to the partial
integration. They also imply $\Diamond \{M(1,2)\}=0$, if $M$ contains no
internal vertices. The convolution
of two two-point functions gives\\[-10mm]
\begin{eqnarray}\label{Poisson}
\hspace*{-5mm}\Diamond \{C(X,p)\}&=&\Diamond 
\left\{\vphantom{\int}\right.\!\DRhomb{A}{B}\!\left.\vphantom{\int}\right\}\cr
&=&
\DPoisson
+\DRhomb{$A$}{$\Diamond B$} +  \DRhomb{$\Diamond A$}{$B$}
\\[1mm]
&=&
\Pbr{A(X,p),B(X,p)}
+A(X,p)\Diamond\{B(X,p)\} +\Diamond\{A(X,p)\}B(X,P),\\
\mbox{where}\quad &&C(1,2)=\oint \di 3 A(1,3)B(3,2).\nonumber
\end{eqnarray}
Besides the Poisson bracket expression $\Pbr{A,B}$ it leads to further
gradients within each of the two functions. Applied to the r.h.s. of
Eq. (\ref{Grad-KB-Eq}), this rule indeed provides the decomposition
into Poisson bracket and memory terms of the quantum kinetic equation
(\ref{Grad-Sep-KB-Eq}).

\end{fmffile}%

\end{document}